\documentstyle[aps,psfig,preprint]{revtex}
\begin{document}




\draft

\title{High Level Trigger Using ALICE ITS Detector}

\author{A. K. Mohanty}
\address{CERN, CH-1211, Geneva 23, Switzerland}

\maketitle

\begin{abstract}
The high $P_T$ trigger capabilities of the ALICE inner tracking system (ITS)
as a standalone detector have been investigated. Since the high $P_T$ charged
particles mostly lead to the linear trajectories within this ITS sector, it is
possible to select tracks of 
$P_T$ of the order of $2$ GeV and above by confining to a
narrow search window in the ($\theta,\phi$) space. Also shown that by 
performing a principal component transformation, it is possible to rotate
from a $12$ dimensional ($\theta$-$\phi$) space 
(in this space, a good ITS track has
$6$ pairs of hit co-ordinates) into a parametric space characterized by
only two independent components when the track momentum exceeds a particular
limit. 
This independent component analysis (ICA) has
been uitilised further to reduce the false track contribution to an acceptable
level particularly when the charged multiplicity is large. 
Finally, it is shown that with  a narrow
bin width of $\Delta \theta = \Delta \phi \sim 0.008$ radian and  with 
 PCA or ICA cut, the ITS can be used to trigger the jet particles
with $P_T \ge 2 $ GeV. Apart from triggering these high $P_T$ particles, this
method can also be used to estimate the initial
momentum of the high $P_T$ tracks for seeding 
which can be further prolonged into the 
TPC detector both for offline and online Kalman tracking or even to
detect those
high $P_T$ tracks of rare events which might get lost in the TPC-TRD dead zone.
 
\end{abstract}



\section {Introduction}

ALICE (A Large ION Collider Experiment) has a dedicated detector systems to study
the collective properties of the hot and dense matter created in nucleus-nucleus
collisions using the CERN large hadron collider (LHC) in the year $\sim 2007$.
The ALICE detectors will measure and identify mid-rapidity hadrons, leptons and photons
produced in the interactions \cite{ppr}. The design parameters of the ALICE detectors
are optimized to cope with multiplicities up to $8000$ charged particles per
unity rapidity (the theoretical prediction of $dN_{ch}/d\eta$ at mid-rapidity 
for PbPb collison at
$\sqrt{s}=5.5$ A TeV may lie between $2000$ and $6000$),
resulting in $20000$ charged primary and secondary tracks in the
acceptance of the central detector systems that covers mid-rapidity $-0.9 \le
\eta \le 0.9$ and full azimuth. Based on current physics simulations, the central
event size in Pb-Pb interaction is expected to be about $86$ MB before compression
out of which
about $75$ MB of raw data is generated by the time projection chamber (TPC),
which is the main tracking device of the ALICE detector (A brief description
of the ALICE detector systems is given in the appendix and also can be found in
ALICE PPR \cite{ppr} and reference therein). Several data compression algorithms
have been tried on simulated TPC data \cite{nic}. 
Using Huffman compression and
an optimized frequency distributioni we have also shown that
 it is possible to achieve a lossless compression ratio
of $\sim 50\%$ on the realistic TPC rawdata that includes read out overheads
\cite{akm1,akm2}. An estimate of the total bandwidth needed for the data transfer
requires evaluation of the rawdata throughput for each type of trigger. The number of
events required to accumulate enough statistics in one year period of data taking
(which has an effective time of $10^6~s$) is a few $10^6$ events for hadronic physics,
a few $10^7$ events for minimum bias, hadronic charm and electron physics and at least 
$10^9$ for dimuon physics. The needed rates are then of the order of a few Hz for hadronic
physics, a few tens of Hz for minimum bias, hadronic charm and elctrons and a few hundreds
of Hz for dimuons for the L2 triggers. At a read out frequency of $200$ Hz, a total band
width of $\sim 15$ GB/s coming out of the detector has to be handled 
(This rate 
roughly coincides with the TPC limitation 
\footnote{At an average luminosity
of $10^{27}~cm^{-2}~s^{-1}$, the minimum bias rate (assuming $8$ barn total
cross section) for PbPb collison is $8$ KHz out of which $1$ KHz may be 
considered as central events. At this event rate, ($f=8000$ events/s), and with 
drift time $\tau=90~\mu s$, the fraction
of PbPb double events in the TPC is $[1-exp(-2\tau_{drift}~f)]=0.76$, where
$2\tau$ accounts for the past and present memory of the TPC. The clean minimum bias
PbPb events thus reduces to about $1900$ Hz and the central rate to about $240$ Hz.
}). 

Thus, the amount of data
which needs to be written to mass storage will be about a few PBytes per year. One of the
major objective of the ALICE experiment is to measure rare processes such as jet transverse
energy spectra up to $E_T \sim 200$ GeV and the pattern of medium induced modifications
of charmonium and bottomium bound states. These low cross section rare events (which range down to
one in $10^7$ pp and to one in $10^4$ PbPb events), require full exploitation of all the
available luminosity in order to attain the desired event statistics 
and will generate data volume which will far
exceed what can be tarnsferred to the permanent storage system. ALICE would not be able
to acquire sufficient statistics for many rare probes without rejecting many 
unwanted events by sharpening the momentum cut through a high level trigger (HLT)
mechanism.
Therefore, the HLT system is one of the important aspect of the ALICE detectors which controls
the data flow between the front end electronics of the detectors and the event builder
of the data acquisition system. Using simulated rawdata, all the present HLT studies
examine the possibility of having a fast cluster finder (using
techniques like  Hough transformation etc) leading to a fast pattern recognition
algorithm primarily for the tracks in the TPC sectors \cite{hlt1,hlt2}. In this work,
we consider the possibility of generating 
a high $P_T$ trigger using the hit informations only
from the ITS sectors of the ALICE detectors
where all high $P_T$ charged particle tarjectories are mostly
linear. As will be shown later, using a simple road-finder algorithm 
based on a  narrow   
($\theta,\phi$) search
window, it has been possible to reject significant amount of low $P_T$ backgrounds.
It has been possible to extract
the momentum of these tracklets through a $\chi^2$ minimization procedure
which agree well (within $20\%$) with the actual 
momentum. The computation is also fast enough as the 
track density is very low within a small $\Delta \theta = \Delta \phi$ bin of $\sim 0.008$
radian, thus reducing the number of combinatorics. Further, it is  found that with proper
choice of the  input track co-ordiates (A track is 
represented by a 12-dimensional hit vector in $\theta, \phi$ space),
a principal component transformation can be carried out which
leads to a parametric space with only two independent
variables when the track momentum becomes large so that all the trajectories
are nearly linear. 
This aspect has been utilised further
to reject the false tracks particularly when the track density is very high. It is shown
in this work that this simple road finder with a PCA or ICA cut can be used to reject
low momentum tracks and select 
$P_T \ge 2$ GeV quite efficiently with less computational efforts. This has been 
demonstrated using Hijing generator with jet options \cite{ppr}. In addition, 
since the extracted momentum
is close to the true value within $20\%$, it can also be used as a seed finder for
high $P_T$ tracks which can be further prolonged into the TPC using Kalman filter 
technique or 
even,  ITS as a stand-alone detector, can also be used to know the origin of a 
 high $P_T$ track
which might otherwsise gets lost in the TPC-TRD dead zones. 

\section {ITS detector and ALIROOT simulation} 

The Inner Tracking System (ITS) consists of $6$ cylindrical layers of silicon
detectors, located at radii, $r=4$, $7$, $15$, $24$, $39$ and $44$ cm. It
covers the rapidity range of $|\eta|< 0.9$ for all vertices located within
the length of the interaction diamond $(\pm 1 \sigma)$, i.e. $10.6$ cm along
the beam direction. The number, position and segmentation of the layers are
optimized for efficient track finding and high impact parameter resolution.
In particular, the outer radius is determined by the necessity to match tracks
with those from the Time Projection Chamber (TPC), and the inner radius is the
minimum allowed by the radius of the beam pipe ($3$cm). The first layer has a 
more extended coverage $(|\eta| < 1.98)$ to provide, together with the Forward
Multiplicity Detector (FMD), a continuous coverage in rapidity for the measurement
of charged particles multiplicity. Because of high particle density, pixel
detectors have been chosen for the inner most two layers, and silicon drift 
detectors for the following two layers. The outer two layers have double sided
silicon microstrip detectors. With the exception of inner most pixel layers,
all others layers will also provide energy loss information in the non-relativistic
region. Thus, the ITS has a stand-alone capability as a low $P_T$ particle
spectrometer.

The simulation is carried out using AliRoot, the standard ALICE simulation
and reconstruction package based on the ROOT  object oriented data
analysis framework \cite{aliroot}. 
We use Hijing generator for background simulation. Although, analysis is
confined only to the ITS sector,
we include all the ALICE detectors
and passive materials in AliRoot simulation with the option of magnetic
filed $B=0.5~T$ in the solenoidal region. In the present work, we use a fast
simulator to get reconstructed points (recpoints) from the hits. We also
neglect the vertx smearing and set vertices to $(0,0,0)$ for all the tracks
in the AliRoot simulation.  
We consider only those tracks which have hits in all the six ITS layers.
To minimize the dispersion, we begin with all the reconstructed points
 in the $4^{th}$ layer
with a given ($\theta, \phi$) 
and look for the corresponding points in the remaining five layers within a search
window of $\theta \pm \Delta \theta$ and $\phi \pm \Delta \phi$. The 
$(\theta, \phi)$ co-ordinate in the
$4^{th}$ layer is considered to be the seed for a good track candidate if
it has entries in all the follwing and 
subsequent layers within that search window. Therefore, the
efficiency ($\epsilon$) is defined as the ratio of good tracks 
(as defined above) divided by the total number of seeds or the reconstructed points in the 
$4^{th}$ layer. Since, we have considered only those tracks which have entries
in all six layers, the total number of rec-points in the $4^{th}$ layer 
also corresponds to the total number of tracks or trcaklets that pass through
the ITS.

\begin{figure}
\centerline{\hbox{
\psfig{figure=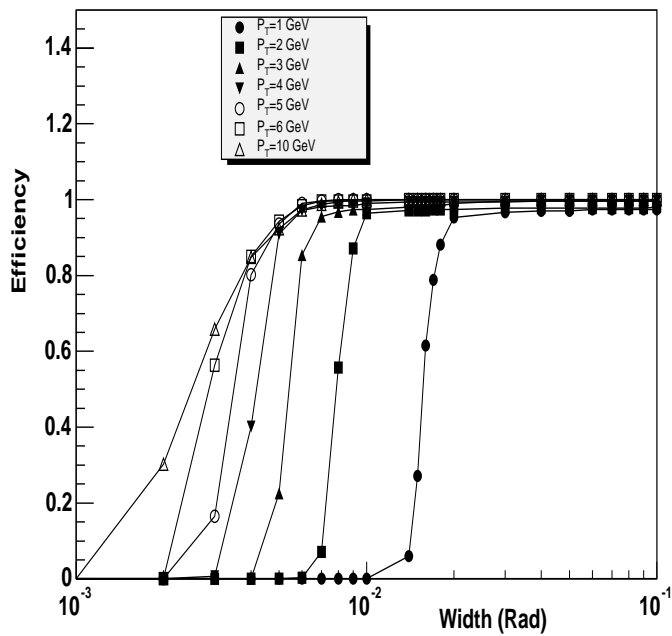,width=4.0in,height=4.0in}}}
\caption{The efficiency $\epsilon$ as a function of bin width ($\Delta \theta
=\Delta \phi$) in radian at different $P_T$.
For detail see the text.}
\end{figure}

Figure 1 shows the efficiency $\epsilon$ as a function of $(\theta,\phi)$
bin width  for different $P_T$. 
As can be seen
from the figure, at a bin width of $\sim 0.008$ radian, it is possible to reject most
of the low $P_T$ tracks below $1.5$ or $2.0$ GeV. For this plot, we have
considered only $2000$ charged pions generated randomly at a fixed $P_T$ within
the full ITS acceptance. 
We have also taken $\Delta \phi=\Delta \theta$, although the spread
in $\theta$ direction is slightly smaller than the $\phi$ dispersion and can 
be optimized further. However,
the choice of $\Delta \phi = \Delta \theta$ only increases the number of combinatorics
marginally and does not affect the analysis. 
Within this ($\theta,\phi$) window, we consider all the combinatorics and extract
the radius of curvature ($R$) by a simple curvature fitting and by accepting the one
for which the $\chi^2$ is minimum. Finally, we estimated the momentum (in GeV)
from the relation $P_T=.002998~B~R$.

To see how well we can reconstruct the transverse momentum by only a $\chi^2$
minimization, we generate charged pions between $0$ to $10$ GeV. However, this
time, we chose $N=4800$, but restrict to only $|\eta| < 0.3$ over full azimuth
so that the 
data volume is not too high and also the number of
primaries chosen corresponds to $dN/d\eta \approx 8000$.
If the fitted value of $P_T$ is within
$20\%$ of the true $P_T$, we consider the estimate as true. Otherwise, it is
a false estimate although the estimated $\chi^2$ is still a minimum. 
Figure 2 shows the plot of number of true and false tracks as defined above as
a function of $P_T$ for three different bin widths of $0.008$, $0.010$ and $0.015$
radians respectively. Although the original $P_T$ distribution is uniform,
the extracted momentum distribution shows non-uniformity due to poor 
momentum resolution.
Note that for $P_T > 2$ GeV, the contribution of the false tracks remain
below $10\%$.  However, it is not possible to extract $P_T$ below $2$ GeV as the false estimate
increases sharply particularly when bin width increases. It is also noticed that
the contributions of the true estimates depend on the bin width to
some extent. For larger bin width, the
contribution of the true estimate decrases while the false estimate goes up
at low momentum below $P_T < 2$ GeV. 
Similarly, if
the bin width is too small, some of the high $P_T$ tracks may get rejected.
Therefore, it is very crucial to use proper bin width.
However, as will be shown below, the dependency on the bin width can be reduced
by incorporating additional techniques like principal component or independent
component analysis.

\begin{figure}
\centerline{\hbox{
\psfig{figure=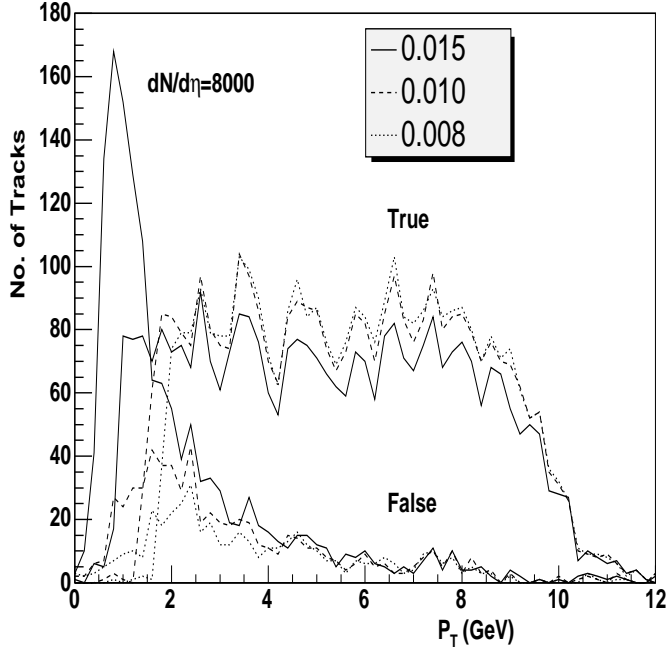,width=4.0in,height=4.0in}}}
\caption{The estimated true and false tracks as a function of $P_T$.
The true tracks are having both a minimum $\chi^2$ values as well as
the extracted $P_T$ values are within $20\%$ of the true transverse momentum.
The false tracks, although have a minimum $\chi^2$, do not give correct
$P_T$ estimate.}
\end{figure}

\section {Independent Component as a High Pass Filter:}
For multivariate data analysis, suitable representation is often sought
as a linear transformation of the original data into a lower dimensional parametric
space. Well known linear transformation methods are principal component analysis
(PCA), factor analysis, and projection pursuit. Independent component
analysis (ICA) is relatively a recently developed method in which the goal is to
find a linear representation of non-Gaussian data so that the components are
statistically independent, or as independent as possible. Such a representation
seems to capture the essential structure of the data in many applications, including
feature extraction and signal separation \cite{ind}. Basically, two random variables
$y_1$ and $y_2$ are said to be independent if information on the value of $y_1$
does not give any information on the value of $y_2$, and vice versa. Mathematically,
if $y_1$ and $y_2$ are said to be independent if and only if the joint 
probability distribution function (pdf) is 
factorizable so that 
\equation p(y_1,y_2)=p_1(y_1)p_2(y_2). \endequation
This definition extends naturally for any number of $n$ random variables, in which
case the joint pdf must be a product of $n$ terms. Therefore, given any
two functions
$f_1$ and $f_2$, the independent random variables should have the property
\equation E[f_1(y_1)f_2(y_2)]=E[f_(y_1)]E[f_2(y_2)], \endequation
where $E$ stands for an  average.
A weaker form of independence is uncorrelatedness. Two random variables are said
to be uncorrelated if their covariance is zero:
\equation E[y_1y_2]=E[y_1]E[y_2]. \endequation
Thus, if the variables are independent, they are also
unclorrelated. On the otherhand, uncorrelatedness does not imply independence
for non-Gaussian random variables. There are many ICA techniques  that lead to
independent estimation of the parameters in the feature space.
Common to all ICA analysis is the PCA transformation which is generally
used as a first step to achieve uncorrelatedness. However, in the present 
application, we will show that by choosing the proper input parameter space, 
it is possible
to achive independence only through a PCA transformation 
although the input variables are known to be non-Gaussian.

The principal component analysis does a simple co-ordinate transformation
to principal axes such that the variances of the new co-ordinates are equal
to the eigen values of the covariance matrix, which is generated out of a large
sample space \cite{pca,dutta}. 
We consider only the ITS layers and a single track
is chracterized by a $M$-dimensional hit vector $h=(h1,h2...,h_M)$ whose average $(H_i)$
and covariance $(A_{ij})$ are
\equation H_i=\frac{1}{N}\sum_{n=1}^N (h_i)_n, \endequation
\equation A_{ij}=\frac{1}{N} \sum_{n=1}^N [(h_i)_n-H_i][(h_j)_n-H_j], \endequation
where $N$ is the number of hit vectors of the $(N \times M)$ dimensional sample space. Consider
the following linear transformation:
\equation X_j=\sum_{i=1}^M \omega_{ij}h_i \endequation
It can be proved \cite{pca} that the variances of $\bf {X}$ are the eigenvalues 
$\lambda$ of the dispersion matrix $\bf{A}$, when $\omega_{ij}$ represents the eigenvectors
of $\bf {A}$. Hence,
\equation Var(X_j)=E\big[ X_j-E\big(\sum_{i=1}^M \omega_{ij}h_i \big)\big]^2 
                  =E(\eta_j^2)=\lambda_j, \endequation
where $E$ stands for average over $N$ and $\eta_j$ is given by
\equation \eta_j=\sum_{i=1}^M \omega_{ij}(h_i-H_i). \endequation
Therefore, the PCA is tuned with a training data set in order to calculate the
average and the dispersion matrices $\bf {H}$ and $\bf {A}$, respectively. The
experiment records a set of $(\theta,\phi)$ pair from each layer when a charged
particle enters the ITS detector. We consider all possible combinations 
$C=(C_1, C_2,...,C_M)$ and estimate the generalized distance:
\equation \eta_j=\sum_{i=1}^M \omega_{ij}(C_i-H_i), \endequation.
\equation d=\frac{1}{M} \sum_{j=1}^M \eta_j^2/\lambda_j. \endequation
It is now required that for a track candidate, $d$ is less than some maximum value
$d_{max}$.

\begin{figure}
\centerline{\hbox{
\psfig{figure=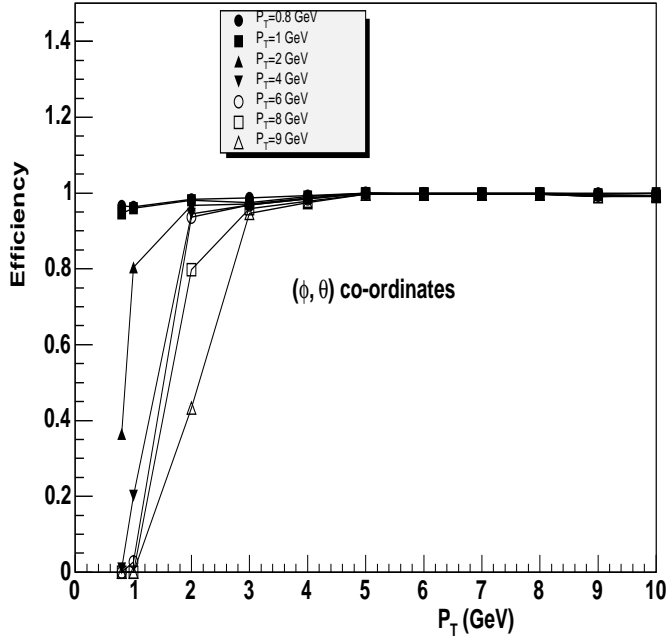,width=4.0in,height=4.0in}}}
\caption{The efficiency as a function of $P_T$.
The good tracks are those for which the generalized
distance $d$ in ($\theta,\phi)$ space is less than $d_{max}=4.0$. The
different curves correspond to training set generated at different $P_T$
as shown in figure caption.}
\end{figure}

In the following, we carry out the PCA test with different training set.
As in figure 1,  we generate $2000$ charged pions at different $P_T$. First,
we consider a hit vector in $(\theta,\phi)$ space for which $M=12$ as a good
ITS track has $6$ pairs of $(\theta, \phi)$ coordiantes. Figure 3 shows
the plot of efficiency of accepting good tracks after a PCA cut as a function
of $P_T$. As in figure 1, here the efficiency is defined as the ratio of the
number of track candidates for which $d \le d_{max}=4.0$ to the total number
of good tracks $N$. The PCA is tuned with  training data set generated 
at different $P_T$.
The symbols in figure 3 show the PCA efficiency when the training set
is chosen at different $P_T$. It is interesting to note that when the PCA
is tuned with the training data set
at $P_T=0.8$ GeV, all track candidates above $P_T > 0.8$ GeV
pass the PCA test (see the filled circle). However, if the PCA is tuned with
a training set at higher momentum, say $P_T=9$ GeV (see open traingles),
then most of the low $P_T$ tracks, do not pass the PCA test.
This is an interesting property which can be utilised to reduce the
tracks of low $P_T$ momentum by appropriate choice of the training data
set.

\begin{figure}
\centerline{\hbox{
\psfig{figure=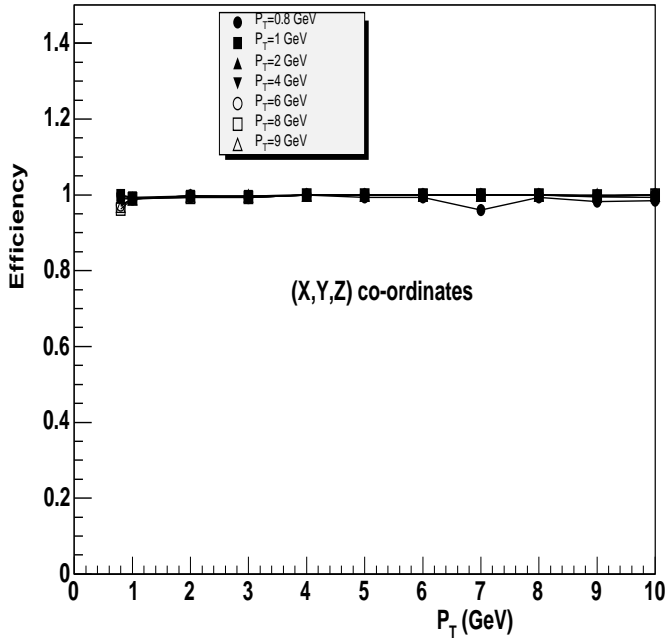,width=4.0in,height=4.0in}}}

\caption{The efficiency as a function of $P_T$.
The good tracks are those for which the generalized
distance $d$ in ($x$, $y$, $z$) space is less than $d_{max}=4.0$.}

\end{figure}

We ahve also carried out similar exercise by considering a hit vector in 
$(x,y,z)$ 
co-ordinates. Figure 4 shows the corresponding plot as that of figure 3. However, with
the choice of the cartesian co-ordinates, PCA does not have any discriminating
power. This aspect can be understood if we examine the number of independent
variables which can be obtained through a PCA transformation. Figure 5 shows the
plot of number of independent variables ( out of $12 ~ \eta^2$, we consider only
those variable whose values are significant and are
 above a given threshold) as a
function of $P_T$. Notice that as $P_T$ increases, the number of independent
variables reduces to two which is a characteristic of all straight trajectories
which can be described by only two independent 
variables. Such behaviour we do not find, if
the input variables are chosen from the cartesian space. Therefore, PCA 
transformation in $(\theta,\phi)$ space is nearly equivalent to an ICA
transformation and can  be used as a high pass filter for
high $P_T$ tracks.

\begin{figure}
\centerline{\hbox{
\psfig{figure=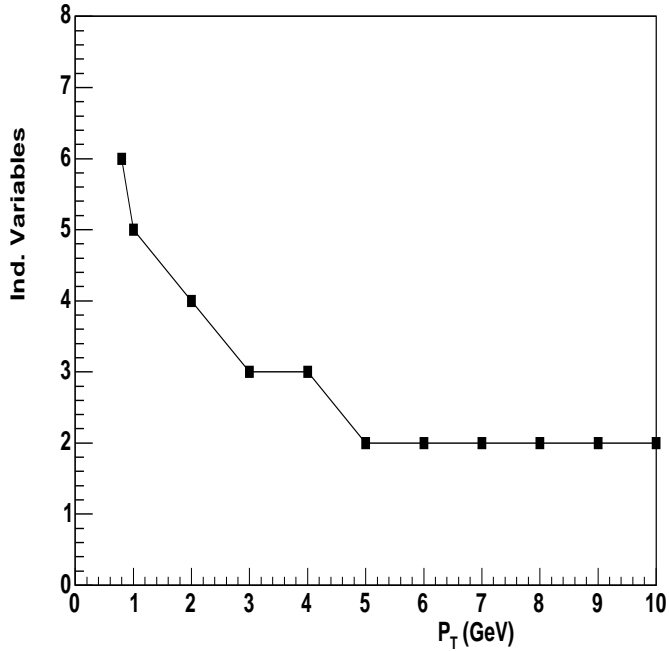,width=4.0in,height=4.0in}}}
\caption{The number of independent variables as a function of $P_T$ ($\theta,\phi$) space.}
\end{figure}

\section {Road Finder with PCA}
In the follwoing, we consider the road finder with the combined action of PCA
and $\chi^2$ minimization. The steps are as follows: (a) First, we consider all
the reconstructed points in the $4^{th}$ layer. Starting with a
given $(\theta, \phi)$,
we pick up all the points
in the remaining $5$ layers within a search window of $\theta \pm \Delta \theta$
and $\phi \pm \Delta \phi$. (b) Second, we consider all the combinatorics and
accept those which qualify a PCA test with $d \le 4.0$. The PCA parameters are
tuned with a training set generated at $P_T=3.0$ GeV \footnote{
We have optimized $P_T$ at $3.0$ GeV for training so that low
$P_T$ tracks below $\sim 2$ GeV can be rejected.}. (c) Next, we extract
the best radius parameter  ($R$) through a minimum $\chi^2$ curvature fitting. 
Therefore,
at this stage, only one combination is retained for which the $\chi^2$ is minimum. 
Finally, the $P_T$ (in GeV) 
is extracted through the relation $P_T=0.002998~B~R$ where 
$B=0.5~T$. As in figure 2, figures $(6-8)$ now show the contributions of true and
false estimates at different track densities both with (steps
$a$, $b$ and $c$) and without PCA cut (steps
$a$ and $c$). With inclusion of PCA, the true estimates become independent of
bin widths as can be seen prominently in figure 8 where the track density is very
high. This is an attractive feature as we need not optimize too much on the bin width
as long as it is close too $0.008$ radian or higher. Although, slight 
higher in bin width increases
the number of combinatorics, it still safer to allow little larger width so that
we do not loose too much low $P_T$ tracks. Another important aspect of the
inclusion of PCA test is that the false $P_T$ 
estimation below $2$ GeV reduces drastically. In short, with a PCA cut ($d \le 4.0$)
and a bin width of $\sim 0.008$ radian, it is possible to extract most of the
high $P_T$ tracks above $2$ GeV with good efficiency.

\begin{figure}
\centerline{\hbox{
\psfig{figure=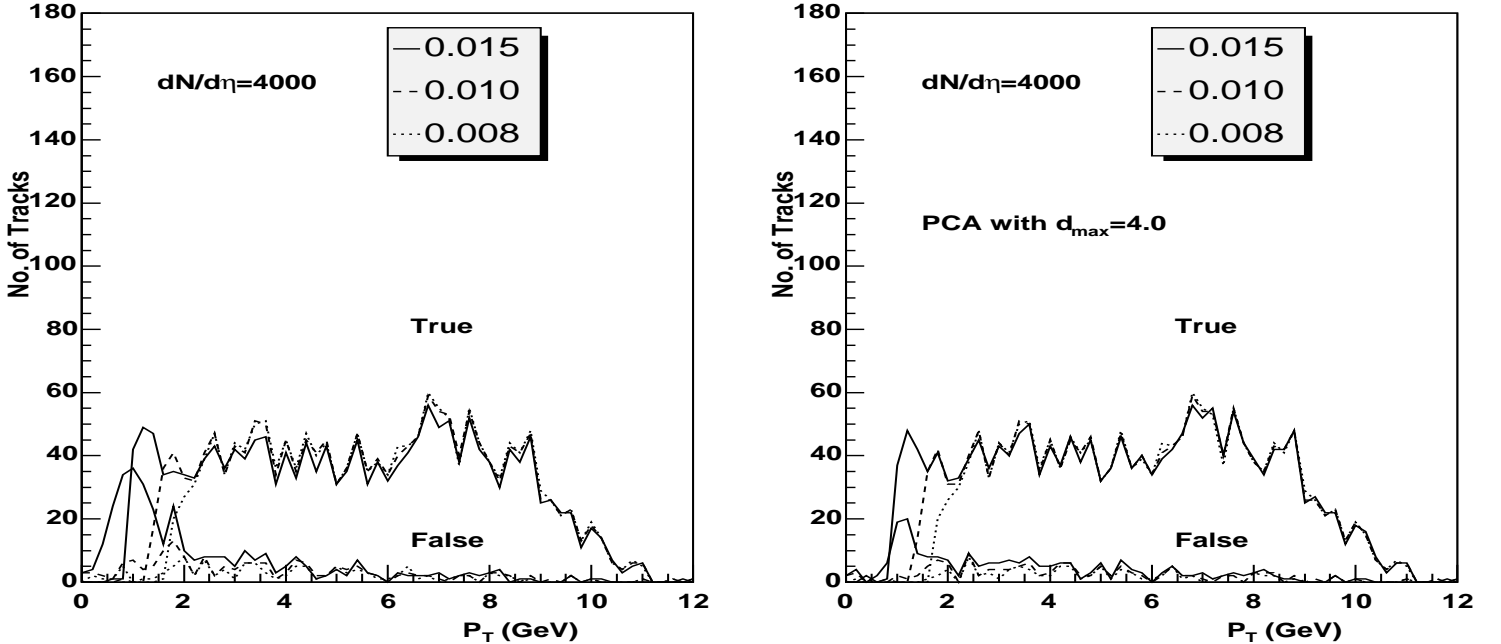,width=8.0in,height=4.0in}}}
\caption{The estimated true and false tracks as a function of $P_T$ both
without (left) and with (right) PCA cut.}
\end{figure}

\begin{figure}
\centerline{\hbox{
\psfig{figure=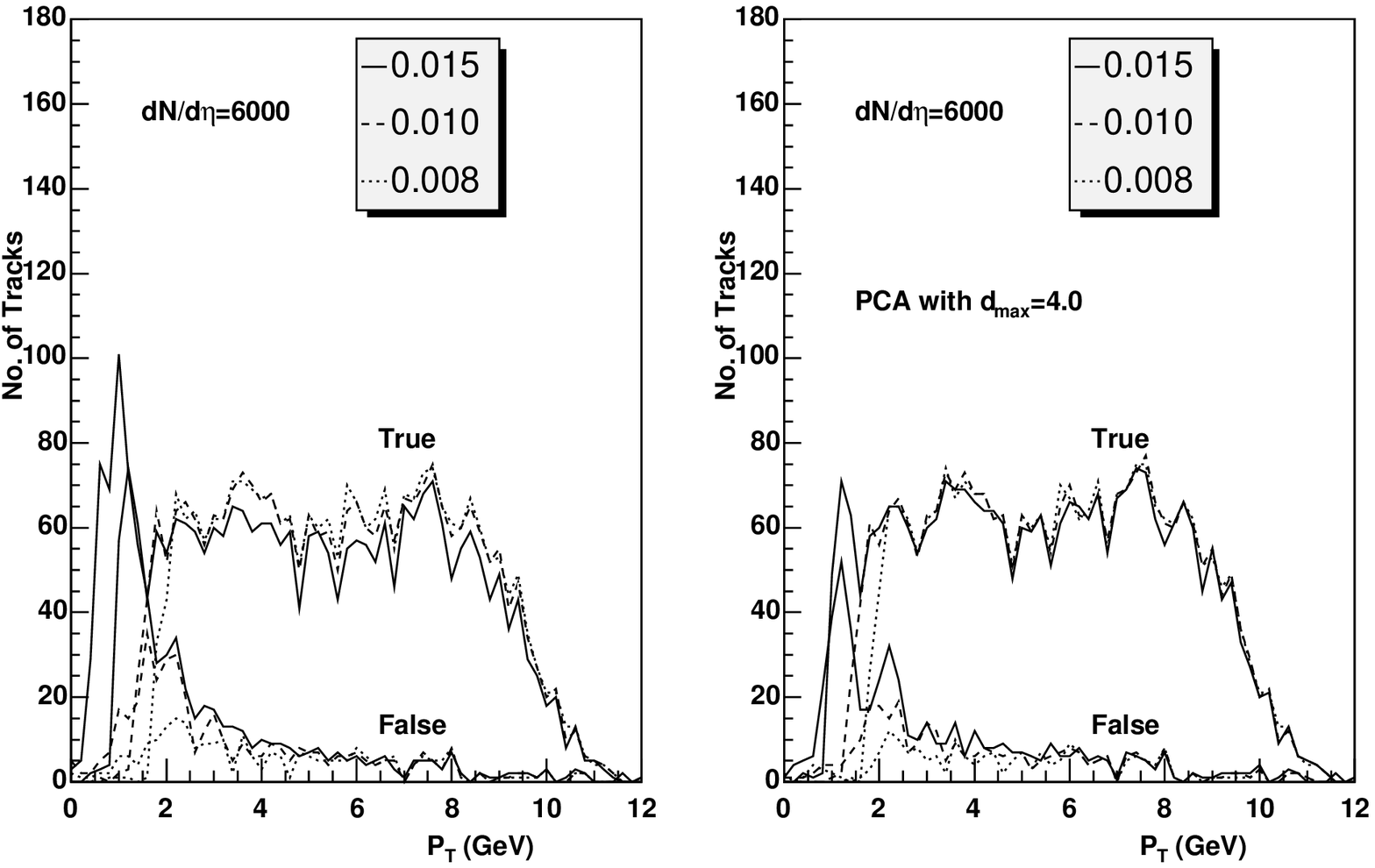,width=8.0in,height=4.0in}}}
\caption{The estimated true and false tracks as a function of $P_T$ both
without (left) and with (right) PCA cut.}
\end{figure}

\begin{figure}
\centerline{\hbox{
\psfig{figure=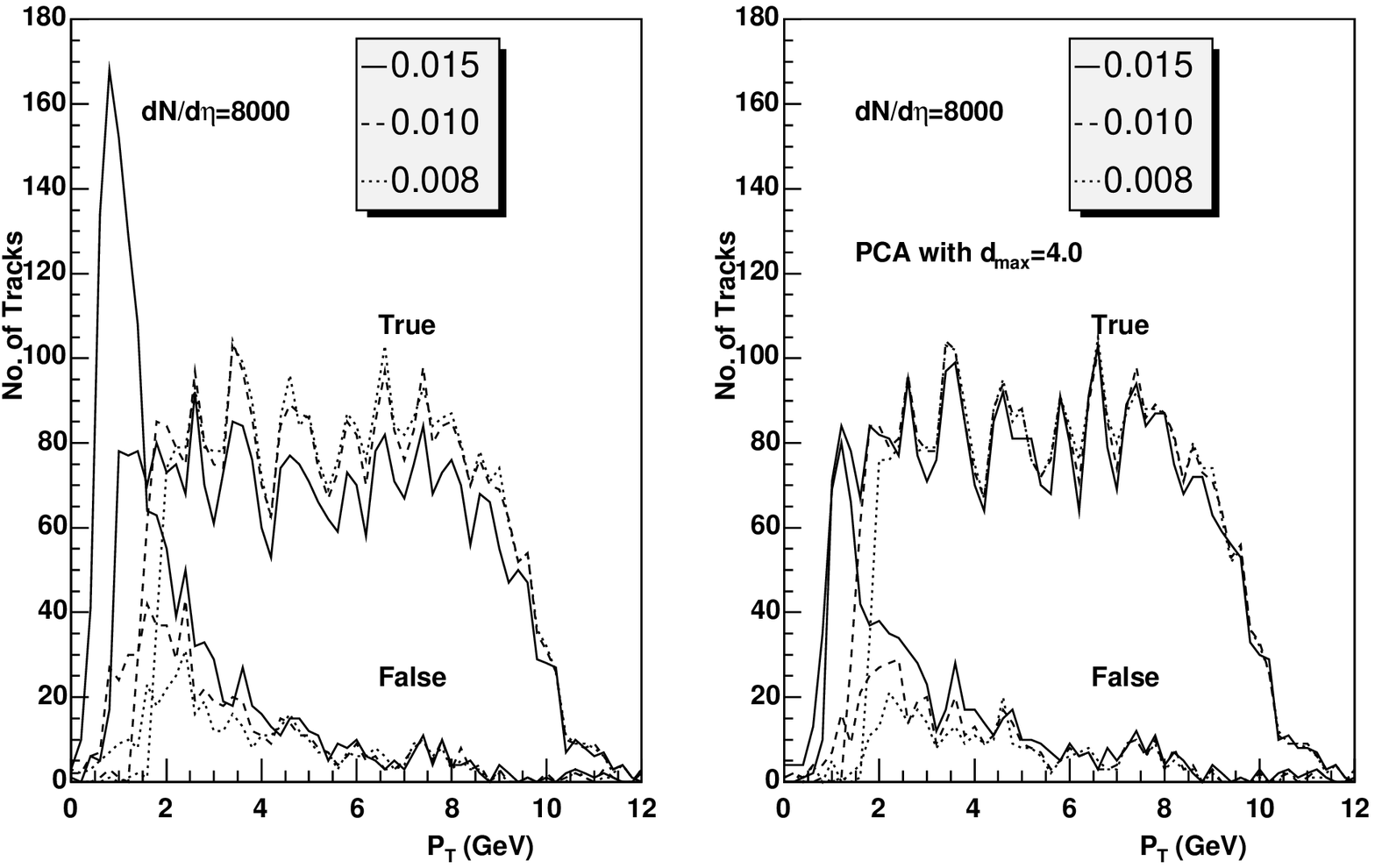,width=8.0in,height=4.0in}}}
\caption{The estimated true and false tracks as a function of $P_T$ both
without (left) and with (right) PCA cut.}
\end{figure}

\section {HLT efficieny with Hijing background}

In the following, we carry out more realistic analysis using Hijing event
generator.
Figure 9 shows the $P_T$ distribution of a
single Hijing event for $|\eta| < 0.3$ over full azimuth 
with jet options.
These $P_T$ distribution
corresponds to the simulated $P_T$ of the rec-points found only at the $4^{th}$ 
layer.  For the present purpose, we may consider it as a respresentation of the
track density distribution inside the ITS. 
Figure 10 shows the $P_T$ distribution
which is obtained after the HLT cut as discussed before with a bin width of
$0.008$ radian and $d_{max}=4.0$. Note that in figure 10, we have only
ploted those tracks having a $P_T > 2$ GeV obtained using the HLT algorithm
as discussed before. These tracks are mostly due to the jet events.

\begin{figure}
\centerline{\hbox{
\psfig{figure=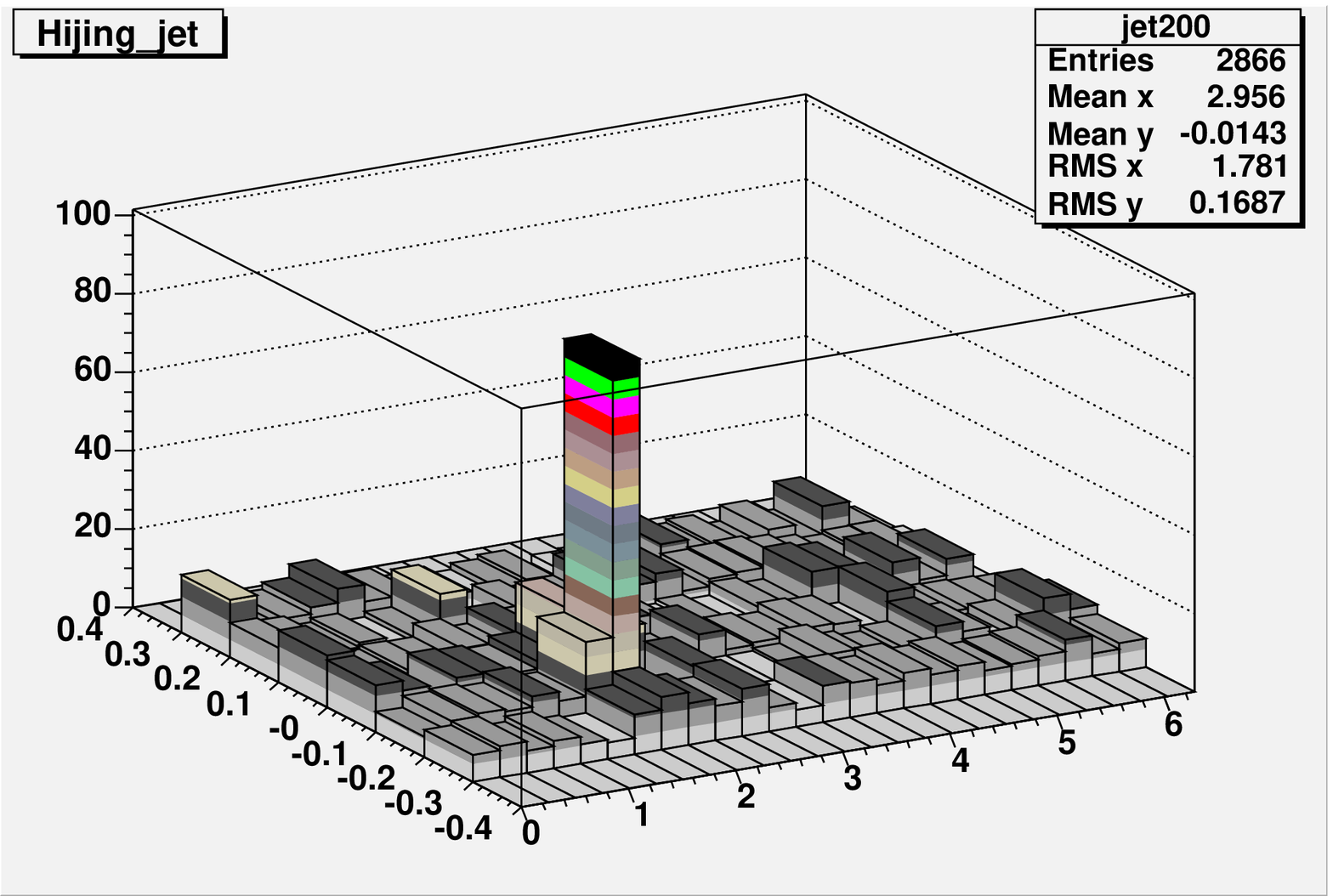,width=4.0in,height=4.0in}}}
\caption{The $P_T$ distribution of recpoints of ITS $4^{th}$ layer
simulated using hijing event generator with jet option.
Here, the bin widths
are $\Delta \eta =0.25$ and $\Delta \theta=0.1$ respectively. This
is the typical jet-cone width. Again, these widths are used only
for bining in the plot. Note that this width is different from the
bin width of $0.008$ radian used in the HLT road finder algorithm. }
\end{figure}

\begin{figure}
\centerline{\hbox{
\psfig{figure=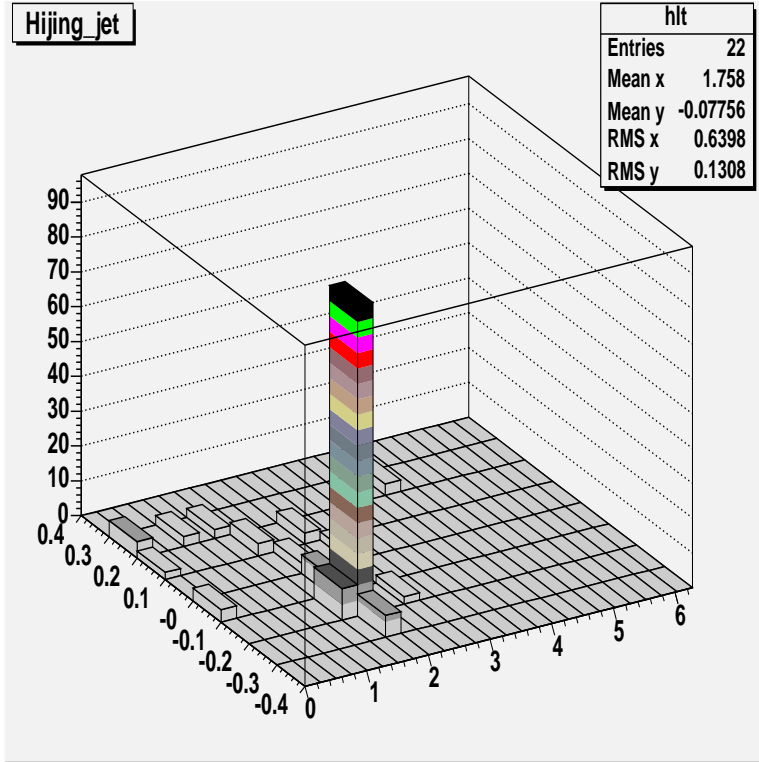,width=4.0in,height=4.0in}}}
\caption{Same as figure 9, but with HLT algorithm}.
\end{figure}

\section {Conclusion}
In this report, it is shown that the Innner Tracking System (ITS) of ALICE
detector can be used for triggering high $P_T$ particles as stand-alone
device. The seed for the trigger road can be selected some where from the middle
layer in order to have minimum dispersion both in $\theta$ and $\phi$ directions.
It is found that most of the background low $P_T$ tracks can be rejected by confining
the search window to a narrow bin width of $\Delta \theta = \Delta \phi=0.008$ 
radian. In order to estimate the transverse momentum, it is required to consider
all the combinatorics inside a given search window which may go up when the
track multiplicity is high. However, it is shown that the number many false
combinatorics can be reduced to an acceptable levels through a PCA test followed
by a least square minimization. Using Hijing event generator, it is shown here
that the ITS can be used to trigger 
charged particles for $P_T$ above $2$ GeV or more. The proposed HLT algorithm
with ITS can be used to trigger high $P_T$ jet particles with good efficiency.

\appendix
\setcounter {figure}{0}
\renewcommand{\thefigure}{A\arabic{figure}}
\section *{A}
The ALICE detector consists of a central detector system, covering
mid-rapidity $(|\eta| \le 0.9)$ over the full azimuth, and several forward systems.
The central system includes, from the interaction vertex to the outside, six
layers of high resolution silicon detectors (Inner Tracking System-ITS), a Time
projection Chamber (TPC)-the main tracking device of the experiment, a transition
radiation detector for electron identification (TRD), and a particle identification
array (Time Of Flight-TOF). The central system which is installed inside a large
soleniodal magnet with field of $\le 0.5~T$ also includes an ring imaging Cherenkov
detector ($|\eta| \le ~0.6$, $56.6^o$ azimuthal coverage) for the identification
of high momentum particle (High Momentum Particle Identification Detector HMPID), 
and an electromagnetic calorimeter ($|\eta| \le 
~ 0.12$, $100^o$ azimuthal coverage) consisting of arrays of high density crystals
(PHOton Spectrometer-PHOS). The large rapidity systems include a muon spectrometer
($-4.0 \le \eta \le -2.4$), a photon counting detector (Photon Multiplicity 
Detector-PMD, on the opposite side), an ensemble of multiplicity detectors (FMD) in the
forward rapidity region (up to $\eta=5.1$). A system of scintillators and quartz counters
($T_0$ and $V_0$) will provide fast trigger signals, and two sets of neutron and hadron
calorimeter, located at $0^o$ and about $90~m$ away from the interaction vertex, will
measure the impact parameter (Zero Degree Calorimeter-ZDC). An absorber positioned
very close to the vertex shields the muon spectrometer. The muon spectrometer consists
of a dipole magnet, five tracking stations, an iron wall (muon filter) to absorb remaining
hadrons, and two trigger stations behind the muon filter.

\end{document}